\documentclass[twocolumn,showpacs,preprintnumbers,amsmath,amssymb]{revtex4}

\usepackage{graphicx}
\usepackage{dcolumn}
\usepackage{bm}

\begin{document}
\preprint{APS/123-QED}

\title{Photoproduction of neutral kaons 
on the liquid deuterium
target
in the threshold region
}

\author{K.~Tsukada}
\author{T.~Takahashi}
\altaffiliation[Present address: ]{Insitute of Particle and Nuclear Studies, High Energy Accelerator Organization (KEK), Tsukuba, 305-0801, Japan}
\author{T.~Watanabe}
\altaffiliation[Present address: ]{Gifu University, Gifu, 501-1193, Japan}
\author{Y.~Fujii}
\author{K.~Futatsukawa}
\author{O.~Hashimoto}
\author{K.~Hirose}
\author{K.~Ito}
\author{S.~Kameoka}
\author{H.~Kanda}
\author{K.~Maeda}
\author{A.~Matsumura}
\author{Y.~Miura}
\author{H.~Miyase}
\author{S.~N.~Nakamura}
\author{H.~Nomura}
\author{K.~Nonaka}
\author{T.~Osaka}
\author{Y.~Okayasu}
\author{H.~Tamura}
\author{H.~Tsubota}
\author{M.~Ukai}
\altaffiliation[Present address: ]{Gifu University, Gifu, 501-1193, Japan}
\author{H.~Yamauchi}
\author{M.~Wakamatsu}
\affiliation{
Department of Physics, Tohoku University, Sendai 980-8578, Japan
}
\author{T.~Ishikawa}
\author{T.~Kinoshita}
\author{F.~Miyahara}
\author{T.~Nakabayashi}
\author{H.~Shimizu}
\author{T.~Tamae}
\author{H.~Yamazaki}
\affiliation{
Laboratory of Nuclear Science, Tohoku University, Sendai 982-0826, Japan
}
\author{A.~Sasaki}
\affiliation{
Department of Electrical and Electronic Engineering, Akita University, Akita, 010-8502, Japan
}
\author{O.~Konno}
\affiliation{
Department of Electrical Engineering, Ichinoseki National College of Technology, Ichinoseki, 021-8511, Japan
}
\author{P.~Byd\v{z}ovsk\'{y}}
\author{M.~Sotona}
\affiliation{Nuclear Physics Institute, 25068, \v{R}e\v{z}, Czech Republic}

\date{\today}

\begin{abstract}
 The photoproduction process of neutral kaons 
 on
 a liquid deuterium target is investigated near the threshold region, $E_\gamma = 0.8$--1.1 GeV.
 $K^0$ events are reconstructed from positive and negative pions,
 and differential cross sections are derived.
 Experimental 
 momentum spectra are compared with those calculated in the spectator model
 using a realistic deuteron wave function.
 Elementary amplitudes as given by recent isobar models
 and a simple phenomenological model are used to study the effect of the new data
 on the angular behavior of the elementary cross section.
 The data favor a backward-peaked angular distribution of the elementary $n(\gamma, K^0)\Lambda$ process,
 which provides additional constraints on current models of kaon photoproduction.
 The present study demonstrates that the $n(\gamma, K^0)\Lambda$ reaction can
 provide key information on  the mechanism of the photoproduction of strangeness.
\end{abstract}

\maketitle
\section{\label{Sect:intro}Introduction}

 Kaon production on a nucleon or a nucleus
 by the electromagnetic interaction
 provides invaluable information on the strangeness production mechanism,
 since the electromagnetic interaction is better understood than the hadronic interaction.
 Studies using real photon and electron beams
 were started in the early 1950s.
 The $p(\gamma,K^+)\Lambda$ reaction was studied from the threshold up to 1.4 GeV
 by measuring the differential cross sections \cite{Anderson:1962,Peck:1964,Groom:1967,Bleckmann:1970,Fujii:1970gn,Feller:1972ph},
 total cross sections \cite{Erbe:1970cq} and $\Lambda$ polarizations \cite{Groom:1967,Fujii:1970gn}.
 An experiment using the polarized target was also made \cite{Althoff:1978qw}.
 The isobar model for kaon photoproduction was developed to reproduce the experimental results in the 1960s \cite{Thom:1966}.

 Since the 1990s,
 new experiments have been carried out using advanced detector systems
 at accelerator facilities
 which provide higher-quality photon or electron beams.
 There are six isospin channels of elementary kaon photoproduction,
 $p(\gamma,K^+)\Lambda$, $p(\gamma,K^+)\Sigma^0$, $p(\gamma,K^0)\Sigma^+$,
 $n(\gamma,K^0)\Lambda$, $n(\gamma,K^0)\Sigma^0$, and $n(\gamma,K^+)\Sigma^-$,
 though kaon production reactions with proton targets have mainly been investigated to date.
 The $p(\gamma,K^+)\Lambda$ reaction has been studied 
 from the threshold region up to a photon energy of 2.95 GeV
 by measuring the differential and total cross sections
 at ELSA/SAPHIR \cite{Glander:2003jw}, JLab/CLAS \cite{Bradford:2005pt} and SPring-8/LEPS \cite{Sumihama:2005er}.
 Polarization transfer in the $p(\vec{\gamma},K^+)\vec{\Lambda}$ reaction
 has been also measured at CLAS \cite{Bradford:2006ba},
 and
 the beam polarization asymmetries for the $p(\vec{\gamma},K^+)\Lambda$ and $p(\vec{\gamma},K^+)\Sigma^0$ reactions
 have been measured at SPring-8/LEPS \cite{Sumihama:2005er}. To date, several theoretical models have been proposed \cite{Adelseck:1990ch,Williams:1992tp,David:1996pi,Bennhold:1989bw,Janssen:2001wk,Mizutani:1998sd,Lee:1999kd}.
 Although a wide variety of observables in elementary kaon photoproduction processes has been investigated,
 no model has successfully explained them all.
 Therefore, further experimental and theoretical studies are eagerly awaited,
 particularly in other isospin channels.
 In particular, the $n(\gamma,K^0)\Lambda$ reaction is expected to greatly help
 clarify the strangeness photoproduction process.
 This reaction has the following features:
 \begin{enumerate}
  \item [{(i)}] The Born term in the t-channel does not contribute, in contrast to reactions with $K^+$ in the final state, because no charge is involved in the reaction.
  \item [{(ii)}] The coupling constants in the u-channel for $\Sigma^0$ and other isovector exchanges have the same value but are of opposite sign to those in the $p(\gamma,K^+)\Lambda$ reaction due to the isospin symmetry.
 \end{enumerate}
 Because of these characteristics,
 the interference among the diagrams in the $K^0\Lambda$ production process is quite different
 from that in the $K^+\Lambda$ process.
 Due to this differences, it was indicated that
 the angular distribution of the $K^0\Lambda$ production process shows a backward peak \cite{Li:1992tz}.
 Therefore, comparison of $K^0$ and $K^+$ photoproduction data,
 especially the angular distribution,
 is very important for the investigation of the strangeness photoproduction mechanism.
 Note, that the absence of electric charge in the $K^0\Lambda$ process (the s-channel Born term is gauge-invariant by itself)
 makes implementation of hadronic form factors easier than in the $K^+\Lambda$ case.
 Besides the features mentioned above,
 the influence of the higher mass resonances can be considered to be small in the present experiment,
 which
 was carried out near the threshold,
 $E_\gamma = 0.8$--1.1 GeV.

 Prior to the present experiment,
 the photoproduction of neutral kaons on
 a carbon target
 near the threshold energy
 was measured for the first time 
 at the Laboratory of Nuclear Science (LNS) of Tohoku University \cite{Watanabe:2006bs}.
 Neutral kaons were detected via the $K^0_S\rightarrow\pi^+\pi^-$ decay channel
 using the Neutral Kaon Spectrometer (NKS).
 The cross sections were compared with those for $K^+$ photoproduction \cite{Yamazaki:1995bc}
 as a function of the incident photon energy.
 It was found that the cross sections for $K^0$ and $K^+$ production on $^{12}$C
 are almost of the same magnitude.
 The momentum spectra of kaons were also compared with simple model calculations
 based on recent isobar models for elementary production processes.
 However, analysis of the elementary reaction was limited because of the uncertainty
 due to the complex many-body nature of the process on the carbon target.

 As a next step,
 we
 carried out
 measurements of neutral kaons
 with NKS using a liquid deuterium target
 and report the results in this paper.
 In Sec.~\ref{Sect:exp},
 the experimental apparatus and methods are described.
 In Sec.~\ref{Sect:ana},
 event selection,
 derivation of the $K^0$ momentum distributions,
 estimation of the background and detection efficiencies are discussed.
 Section~\ref{Sect:result} presents
 results for momentum spectra for the deuteron target
 and their comparison with theoretical predictions.
 The conclusion is given in Sec.~\ref{Sect:sum}.

\section{\label{Sect:exp}Experimental apparatus}

The $K^0$ meson was inclusively measured
via the $\pi^+\pi^-$ decay channel
using NKS,
which was constructed and installed
in the second experimental hall
equipped with the STB-Tagger system \cite{Yamazaki:2005}
at LNS.
NKS was originally used as the TAGX spectrometer
at the electron synchrotron of the Institute of Nuclear Study, University of Tokyo (INS-ES) \cite{Maruyama:1995ip}.
In this section,
the apparatuses of the beam line and the detectors of NKS are described.

\subsection{\label{Sect:beamline}Tagged-photon beams}
\begin{figure}
 \includegraphics[width=9cm]{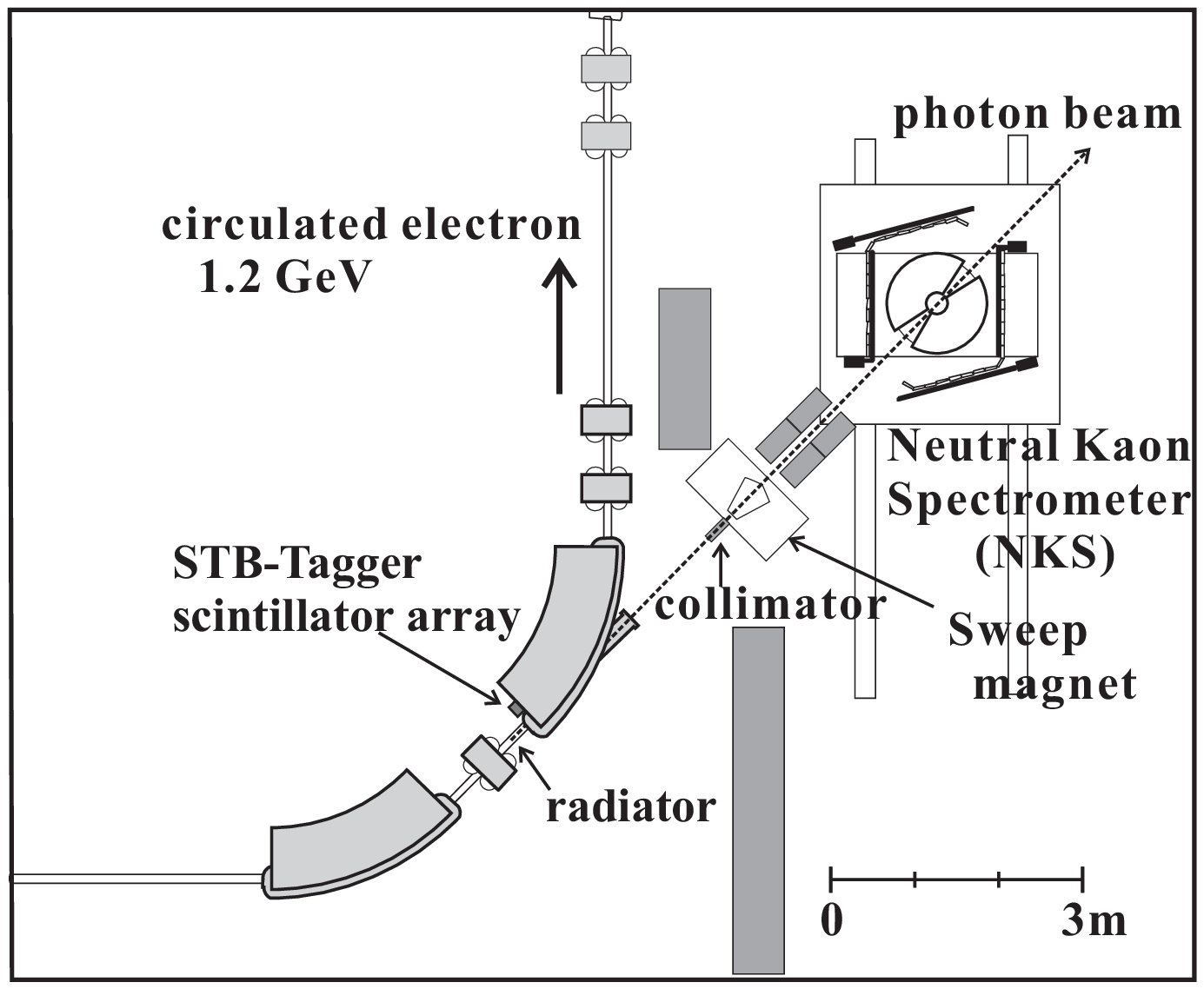}
 \caption{\label{Fig:lns}
 Top view of the experimental setup in the second experimental hall at LNS.
 The electrons were accelerated up to 1.2 GeV in the accelerator ring.
 Bremsstrahlung photons were generated by inserting a radiator.
 The scattered electrons were detected by STB-Tagger.
 A collimator and sweep magnet were set on the photon beam line
 to remove the beam halo and the background from $e^+e^-$ pair creation.
 }
\end{figure}

Figure~\ref{Fig:lns} shows the experimental setup.
A 200-MeV electron beam from the linear accelerator was injected into the STretcher Booster (STB) ring,
accelerated up to 1.2 GeV typically in 1.2 s,
and then stored for 20 s in the ring.
Bremsstrahlung photons were generated by inserting a radiator
made of a carbon fiber of 11 $\mu$m~$\phi$
at
the entrance of one of the bending magnets.
The generated photon energy was tagged by
analyzing the momentum of a scattered electron with the bending magnet \cite{Yamazaki:2005}.
The tagging counter consisted of 50 plastic scintillators,
the time resolution being evaluated to be 240 ps.
The duty factor was about 64\% with an 18 s beam every 25 s.

The tagged energy range was 0.8--1.1 GeV and the accuracy of the energy was estimated to be $\Delta$E = $\pm$10 MeV (see Sect.~\ref{Sect:enecal}).
The total count rate of the counter array was 2 MHz on average.

As shown in Fig. \ref{Fig:lns},
a sweep magnet of about 1.0 T and 0.5 m long 
was located between 
a extraction window
and 
NKS
to suppress the background of the $e^+$ and $e^-$ created
at the exit window of the photon beam.
In addition, a 25 cm long lead collimator with a 1.0 cm diameter aperture was placed just upstream of the sweep magnet to remove the photon beam halo.
Furthermore, two helium bags were installed in the space between the sweep magnet and the target.
The windows of the bags along the beam line were made of mylar film of 10 $\mu$m thickness.
The size of the photon beam at the target was 3--5 mm ($\sigma$).

\subsubsection{\label{Sect:enecal}Energy calibration}

Energy calibration of the STB-Tagger system was carried out
prior to the series of $K^0$ photoproduction experiments with NKS.

In the calibration experiment, 
the sweep magnet was used as a spectrometer for $e^+e^-$ pairs
generated in the $\gamma \rightarrow e^+e^-$ conversion process.
A copper converter of 0.9 mm $\phi$ was installed upstream of the magnet.
The $e^+e^-$ pairs were detected by a drift chamber
installed downstream of the magnet.
The momenta of $e^+e^-$ pairs were analyzed using the magnetic field distribution
calculated by the 3-dimensional finite element method program (TOSCA).
Finally, the photon energy was derived by
$E_\gamma = E_{e^+} + E_{e^-}$.
The precision of the photon energy was determined to be $\pm$ 10 MeV
considering the uncertainty of the magnetic field distribution.

\subsection{\label{Sect:setup}NKS}
\begin{figure}
 \includegraphics[width=9cm]{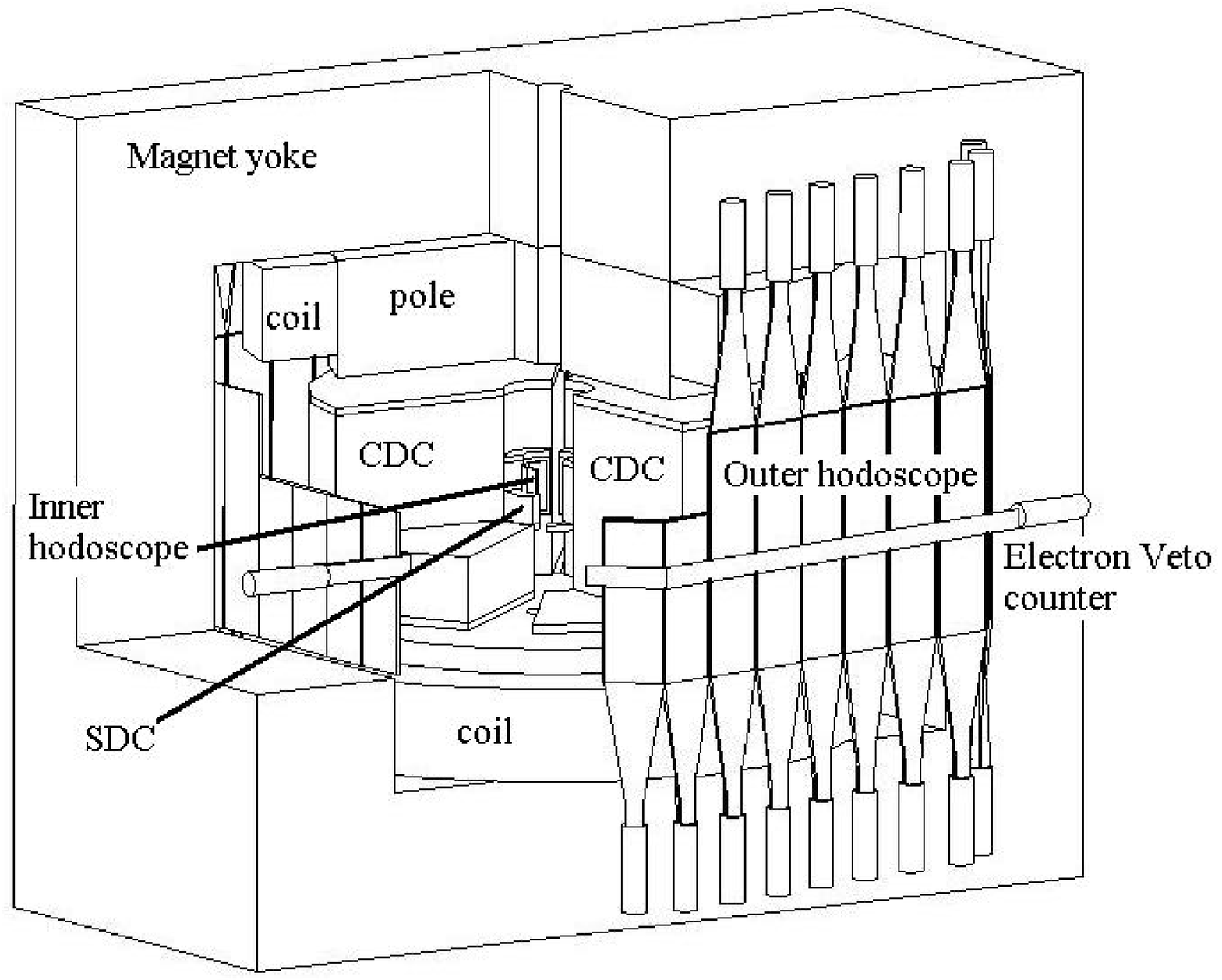}
 \caption{\label{Fig:nks}
 Schematic drawing of NKS.
 The drift chambers (CDC, SDC) were set among the poles.
 }
\end{figure}

\begin{figure*}
 \includegraphics[width=12cm]{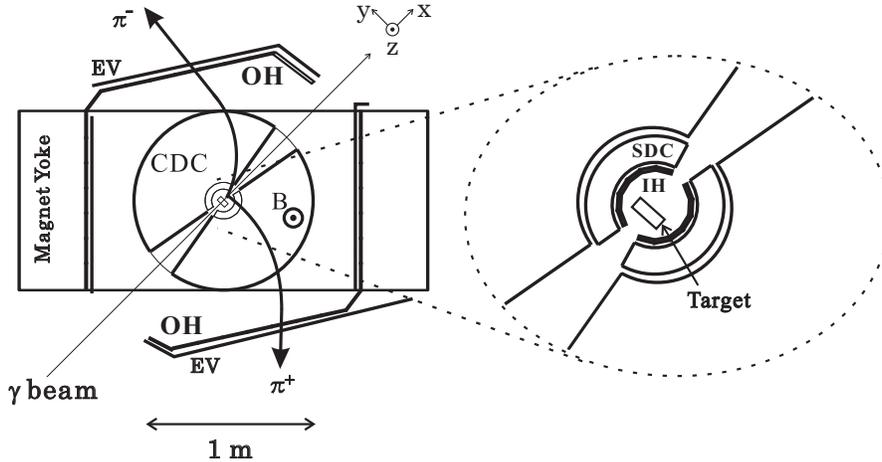}
 \caption{\label{Fig:nks2}
 Floor plan of the detector configuration
 and an example of two pion trajectories.
 The detectors were placed almost symmetrically along the beam line.}
\end{figure*}

Figure~\ref{Fig:nks} shows a schematic drawing of NKS
and Fig.~\ref{Fig:nks2} shows the detector configuration.
The spectrometer consisted of a 0.5-T dipole magnet
of 107 cm diameter and 60 cm gap,
with a pair of cylindrical drift chambers (CDC), a pair of straw drift chambers (SDC), 
inner and outer scintillator hodoscopes (IH, OH) 
and electron veto scintillation counters (EV).

IH
surrounded the target 6 cm from the center.
It was
horizontally
segmented into twelve pieces of 12 cm height and 0.5 cm thickness,
covering the angular range from $\pm15^{\circ}$ to $\pm160^{\circ}$.
The scintillation photons were guided
from the bottom end of each scintillator
to the outside of the magnet gap through bundles of optical fibers about 1.5 m long.
The time resolution of IH was about 470 ps ($\sigma$).

SDC
covered the angular range from $\pm 10^{\circ}$ to $\pm 170^{\circ}$,
and the radial range from 7.18 to 10.19 cm.
SDC had four layers of sense wires
made of gold plated tungsten of 20 $\mu$m~$\phi$
with straw tubes made of aluminized mylar film 180 $\mu$m thick.
CDC
covered the angular range from $\pm 15^{\circ}$ to $\pm 165^{\circ}$,
and the radial range from 13.8 to 48.6 cm.
Field wires were arranged hexagonally around a sense wire.
Since all the wires were vertical,
the information of the track position was horizontal only.
CDC had twelve layers
grouped into four groups, each having three layers.
In the present experiment,
nine layers of the twelve were used since
the third layer of each group was not active.
The CDC sense wires were made of gold plated tungsten of 30 $\mu$m~$\phi$ for the inner two layers
and of stainless steel for the third layers.
The field wires were made of molybdenum of 100 $\mu$m~$\phi$.
The drift chamber gas was a mixture of argon (50\%) and ethane (50\%).
High voltages were applied to the sense wires, typically $+$3000 V for CDC and $+$1900 V for SDC.
The spatial resolutions were about 400 $\mu$m for CDC
and 500 $\mu$m for SDC.

OH hodoscopes were installed outside the drift chambers.
They were
horizontally
segmented into 34 pieces,
the size of each counter being about 60 cm in height, 1.0 cm thick
and typically 15 cm wide.
Some of the scintillators were placed inside the magnet gap
as shown in Fig.~\ref{Fig:nks2},
and
the scintillation photons were guided outside the magnet through the bundle of optical fibers.
All the OH had photomultiplier tubes (PMT) at both ends
and provided information on the time of flight between IH and OH
with a time resolution of 510 ps.
The vertical position of charged particles hitting an OH
was also given by the time difference of the signals from two PMTs.
The vertical position resolution of OH was about 4 cm.

The geometrical acceptance of NKS was about $\pi$ sr
and the momentum threshold was 80 MeV/c.
$K^0$s were identified by the invariant mass of $\pi^+\pi^-$ pair measured
in coincidence with the left and right detector arms.

EV scintillators were installed
on the midplane of NKS and next to the OH array, as shown in Fig.~\ref{Fig:nks2},
in order to reject $e^+e^-$ pairs
which were
distributed
in the midplane
and 
causing large background trigger.
They covered a vertical range $\pm$2.5 cm at the OH array
and therefore the NKS geometrical acceptance was reduced by about 8\%.


\subsection{Liquid deuterium target}
\label{Sect:target}
The liquid deuterium target was inserted through the central
holes of the pole and the yoke of the NKS magnet from the top
of the yoke. The cryostat had a cylindrical part of 1200 mm long
with a diameter of 100 mm for the insertion. The target cell was
contained at the lower end of  the cryostat. The two stage
Gifford-McMahon refrigerator (Sumitomo Heavy Industry RD208B) was
attached at the top of the cryostat. An oxigen free copper rod
with a diameter of 25 mm and a length of 942 mm connected
the second stage and the condenser placed near from the target
cell. The deuterium gas was pre-cooled at the first stage, liquefied
at the consenser, and then stored in the cell.

The target cell was cylindrical and placed with its axis parallel to
the beam line. The length of the cylinder was 30 mm and the
diameter was 50 mm. The cylindrical shell was made of aluminum
of 1 mm thick and  its entrance and exit window were covered with
polyimide (Ube UPILEX-S) films of 75 $\mu$m thick.
The vacuum chamber of the cryostat surrounding the target cell
was made of carbon fiber reinforced plastic (CFRP) of 1.5 mm thick.
It also had a entrance window of 40 mm for the gamma beam covered
with a polyimide film of 75 $\mu$m thick. The target cell was placed at
the 15 mm upstream from the center of NKS. The thickness and
the position of the target was optimized by a simple simulation
for the K0 yield.

The typical temperature of the liquid deuterium was 19 $\pm$ 1 K;
the pressure, 50 $\pm$ 2 kPa; and the density, 0.17 g/cm$^3$ in the data
taking period. The target system was remotely controlled and
monitored with a LabView program running on a Linux-PC for
the easy and secure access via network.


 
 \subsection{Data acquisition system}
 The data for the hodoscope and tagger counters were fed into the TDC and ADC modules of the TKO \cite{Ohska:1985xv}
 , then stored in a VME memory module (SMP)
 and read event by event
 by Linux-PC through a VME-PCI interface module (Bit3-617).
 The signals of the drift chambers were pre-amplified 
 and sent to the amplifier--discriminator cards
 about 7 m away from the chambers.
 The digitized data were transferred by 70 m twisted pair cables 
 to the counting room 
 and were recorded by the Lecroy 4290 TDC system.
 These data were also read by Linux-PC
 through a CAMAC interface (TOYO CC7700).

 A trigger signal was generated when more than two charged particles
 were detected in coincidence with the IH and OH in both the left and right arms,
 and at least one hit in the tagger counters.
 As mentioned before, no EV hits were required.

 The data-acquisition efficiency was typically 90\% with a trigger rate of 100 Hz.

\section{\label{Sect:ana}Analysis and results}
 $K^0$ was identified and measured by $K^0 \rightarrow \pi^+\pi^-$ decay
 in the invariant mass of $\pi^+\pi^-$.
 $\pi^+$($\pi^-$) was identified from the momentum, charge and velocity.
 The horizontal momenta of the charged particles were 
 calculated from the curvature of the trajectories,
 and the vertical momentum information was extracted
 assuming that a vertical projection was a straight line between the spectrometer center and the OH.
 The particle velocities were obtained from the flight length
 and the time-of-flight between the IH and OH.


 \subsection{\label{Sect:ev}Event selections}
\begin{figure}
  \includegraphics[width=8cm]{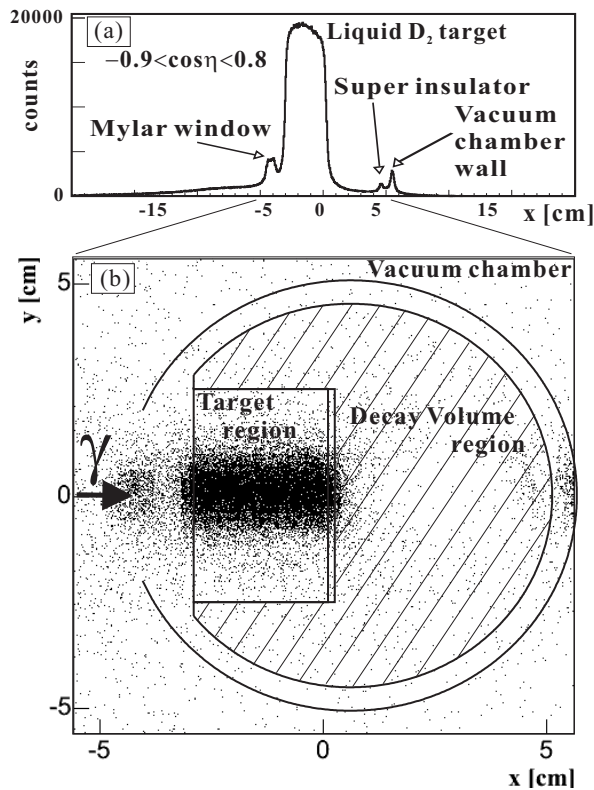}
  \caption{\label{Fig:vert1}
  (a) Vertex point distribution along the beam line for the deuterium target
  after applying a cut of the opening angle ($\eta$).
  (b) Scatter plot of the vertex point distribution around the target.
  The contribution from the walls of the vacuum chamber is also seen.
  }
 \end{figure}

 $e^+e^-$ pair production was the largest source of background
 even using EV in the trigger level.
 In the analysis, $e^+e^-$ pairs generated upstream of the target were removed
 by rejecting upstream vertex points.
 Figure \ref{Fig:vert1}(a) shows a distribution of vertex points
 along the beam line axis.
 Since the acceptance for $e^+e^-$ events generated
 near the target was much smaller than
 for those
 generated in the upstream region,
 most of the $e^+e^-$ events were removed.
 Furthermore, a cut for the opening angle ($\eta$), $-0.9 < \cos\eta < 0.8$, 
 was applied to remove $e^+e^-$ pairs
 because the opening angle of $e^+e^-$ was much smaller than that of $\pi^+\pi^-$ from $K^0$ decay.

\begin{figure}
  \includegraphics[width=9cm]{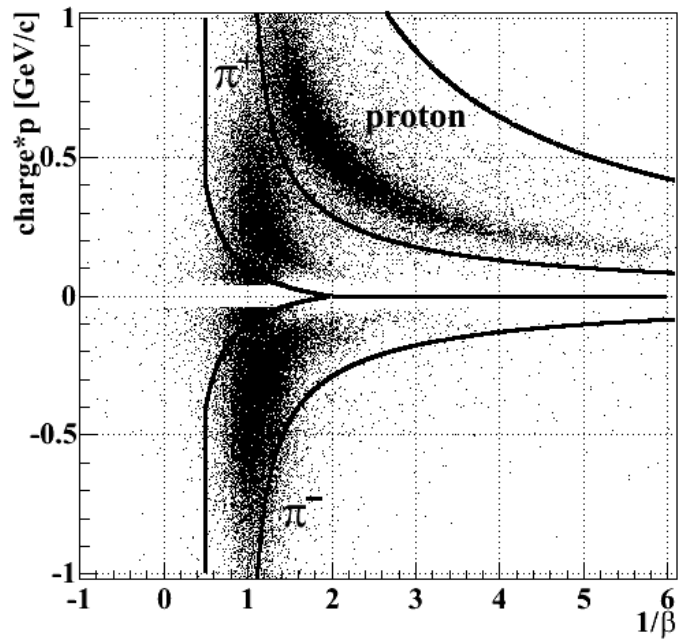}
  \caption{\label{Fig:pid}
  Contour plot of $\beta^{-1}$ vs. momentum.
  The sign of the momentum represents the charge of the particle.
  The solid lines represent the selected region as protons or pions.
  }
 \end{figure}

 Figure \ref{Fig:pid} shows a contour plot of inversed velocities vs. momenta of particles
 after applying the cut of the opening angle
 and selecting decay points near the target.
 The sign of the momentum represents the charge of the particle.
 The particles were identified by selecting the region-defined-lines in Fig.~\ref{Fig:pid}.
 
 Figure \ref{Fig:vert1}(b) shows the vertex point distribution around the target for $\pi^+\pi^-$ events.
 The materials around the target are also indicated in the figure.
 The resolution of the x-position of the vertex was about 1.3 mm.
 
\begin{figure*}
 \includegraphics[width=10cm]{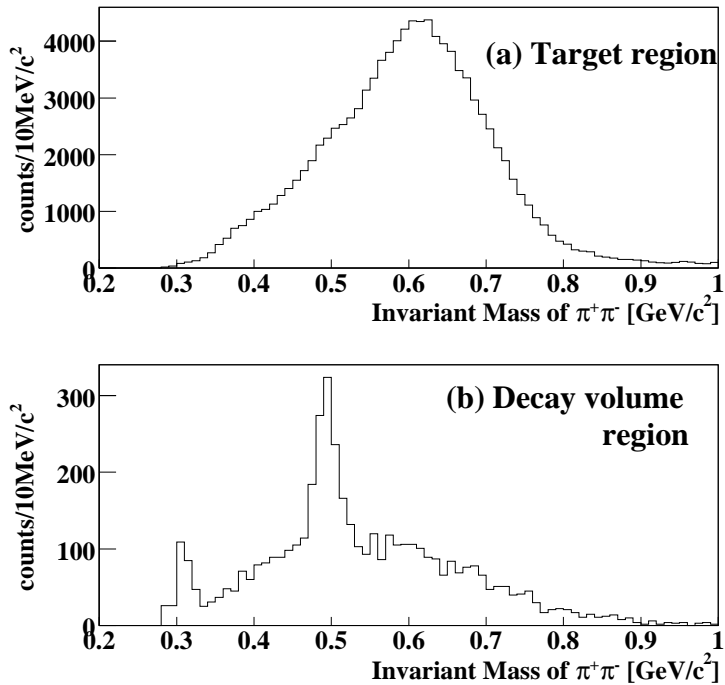}
 \caption{\label{Fig:im}
 Invariant mass spectrum of $\pi^+\pi^-$
 (a) in the target region
 and
 (b) in the decay volume.
 The peak at 0.3 GeV/c$^2$ is caused by low momentum $e^+e^-$ events identified as $\pi^+\pi^-$.
 }
\end{figure*}

 Figure \ref{Fig:im} shows the spectra of the $\pi^+\pi^-$ invariant mass
 for events with vertex points in the target (a) and out of the target (b).
 For events in the target region,
 $K^0$ events were hidden
 in the invariant mass spectrum
 since the spectrum is dominated by processes not involving strangeness, such as production of $\rho$,
 nucleon resonances and others.
 These particles immediately decay by the strong interaction.
 In contrast, for events whose vertex points were reconstructed
 outside of the target,
 a peak of $K^0$ is clearly seen
 because 
 a considerable fraction of $K^{0}_{S}$ decays outside of the target
 due to the relatively long life time of $c\tau \sim$ 2.68 cm.
 Thus, we defined the decay volume as the hatched area shown in Fig.~\ref{Fig:vert1}(b) to select $K^0_S$.
 
\subsection{\label{Sect:background}Background estimation}
\label{Sect:BG}
Even after the background was suppressed as described in the previous section,
a background contribution still exists in the $\pi^+\pi^-$ invariant mass spectrum
as shown in Fig.~\ref{Fig:im}.
The possible origins of the background were considered to come from

\begin{description}
 \item[(1)] imperfect rejection of non-strangeness events
	    such as $\rho$ production
	    due to the finite resolution of the vertex points, and
 \item[(2)] combinatorial background of $\pi^+\pi^-$,
	    such as $\pi^+$ from $K^0_S$ and $\pi^-$ from $\Lambda$.
\end{description}

The shape of background (1) was assumed 
to be the same as that of events 
whose vertex points were reconstructed in the target region
because those events were considered to have almost the same kinematics.
On the other hand, 
the shape of background (2) was evaluated by Geant4 simulation.

\begin{figure*}
 \includegraphics[width=8cm]{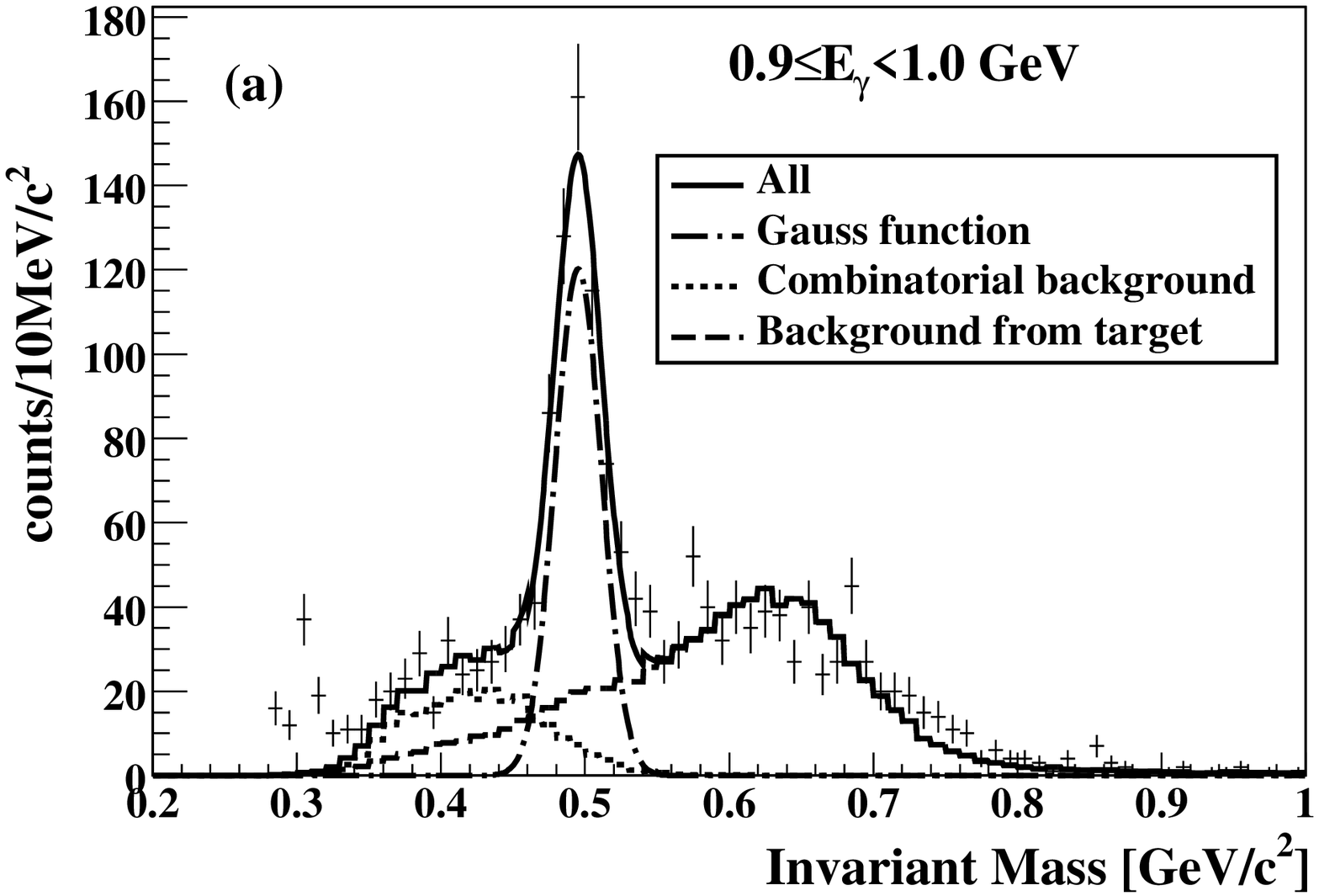}
 \includegraphics[width=8cm]{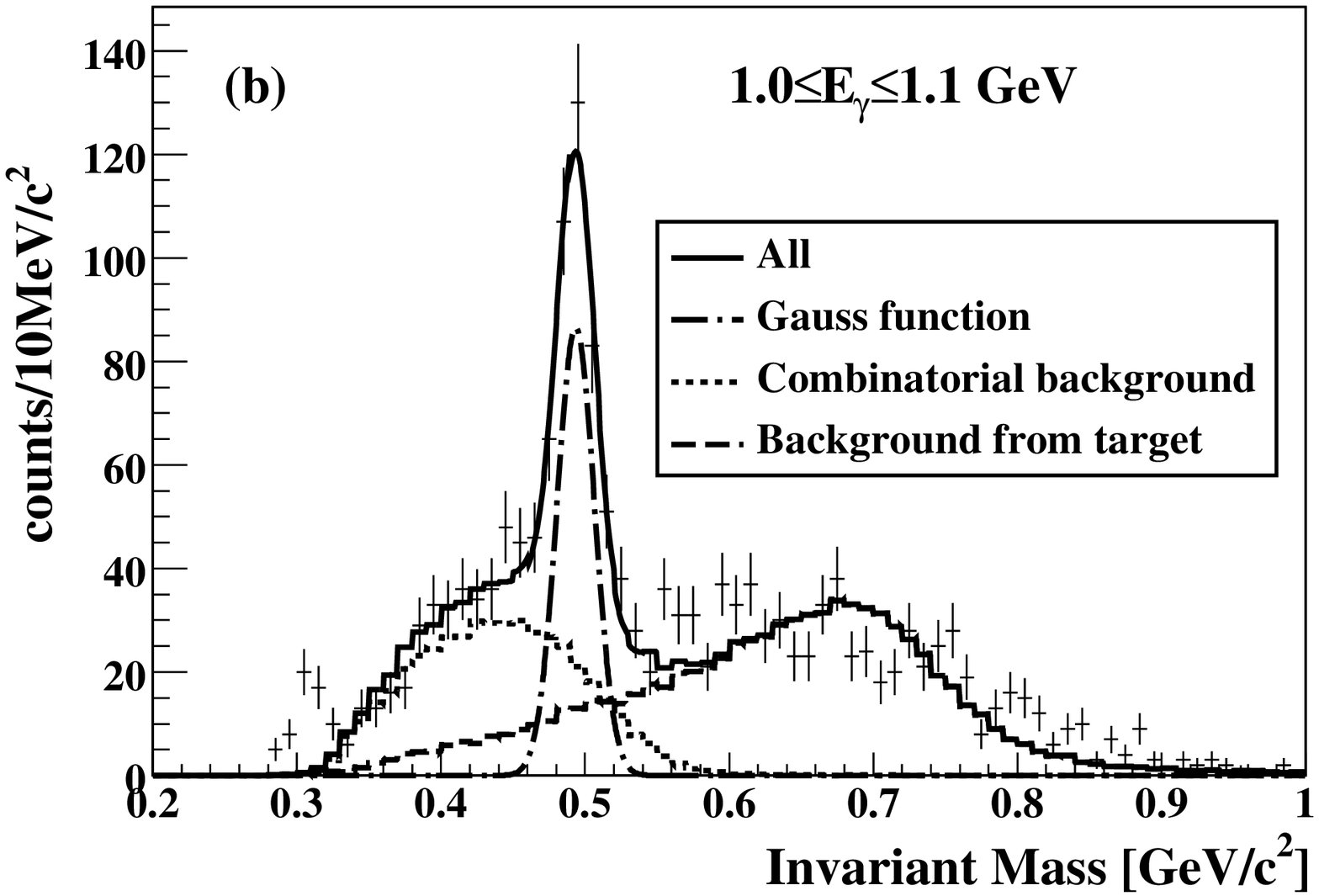}
 \caption{\label{Fig:im-fit}
 Fitting results for $\pi^+\pi^-$ invariant mass spectra in the photon energy ranges (a) from 0.9 to 1.0 GeV
 and (b) from 1.0 to 1.1 GeV.
 The contribution around 0.4 GeV/c$^2$ comes from the combinatorial background
 and that around 0.6 GeV/c$^2$ comes from other process produced in the target region.
 }
\end{figure*}

Figure \ref{Fig:im-fit} shows $\pi^+\pi^-$ invariant mass spectra fitted assuming a Gaussian shape for the $K^0_S$ peak
and the pre-determined shapes for the two backgrounds.
The fitting errors of the background spectra were
typically 4\% for (1) and 8\% for (2).
The widths of the $K^0_S$ peak around 0.5 GeV/c$^2$ were 16.2 $\pm$ 0.8 MeV/c$^2$
for the lower beam energy region ($0.9 \leq E_\gamma < 1.0$ GeV)
and 13.0 $\pm$ 1.4 MeV/c$^2$ 
for the higher beam energy region ($1.0 \leq E_\gamma < 1.1$ GeV).
In further analysis, the mass gate between 0.46 and 0.54 GeV/c$^2$ was applied to select $K^0$ events.

\subsection{\label{Sect:eff}$K^0$ momentum spectra}
The differential cross section of neutral kaons is given by
\begin{widetext}
 \begin{equation*}
  \frac{d^2\sigma}{d\Omega dp} = \frac{N_{yield}(p,\cos\theta)}{N'_{\gamma}\cdot 
   N_{target}\cdot \epsilon_{acpt}(p,\cos\theta)\cdot \epsilon_{DAQ}\cdot \epsilon_{track}
   \cdot \epsilon_{gate} \cdot 2\pi \ \Delta(\cos\theta) \Delta p}\label{Eq:CrossSection},
 \end{equation*}
\end{widetext}
where $N_{yield}$, $N'_\gamma$ and $N_{target}$ are the number of selected $K^0$ events,
the number of photons bombarding the target and the number of target neutrons,
 respectively.
$p$ and $\cos\theta$ are the $K^0$ momentum and the emitted angle to the photon beam line in the laboratory frame,
$\epsilon_{acpt}(p,\cos\theta)$ is the acceptance of NKS for $K^0$ detection, 
$\epsilon_{DAQ}$ is the data taking efficiency,
$\epsilon_{track}$ is the tracking analysis efficiency,
and $\epsilon_{gate}$ is the efficiency due to the gating $K^0$ mass spectrum.

\begin{figure*}
 \includegraphics[width=8cm]{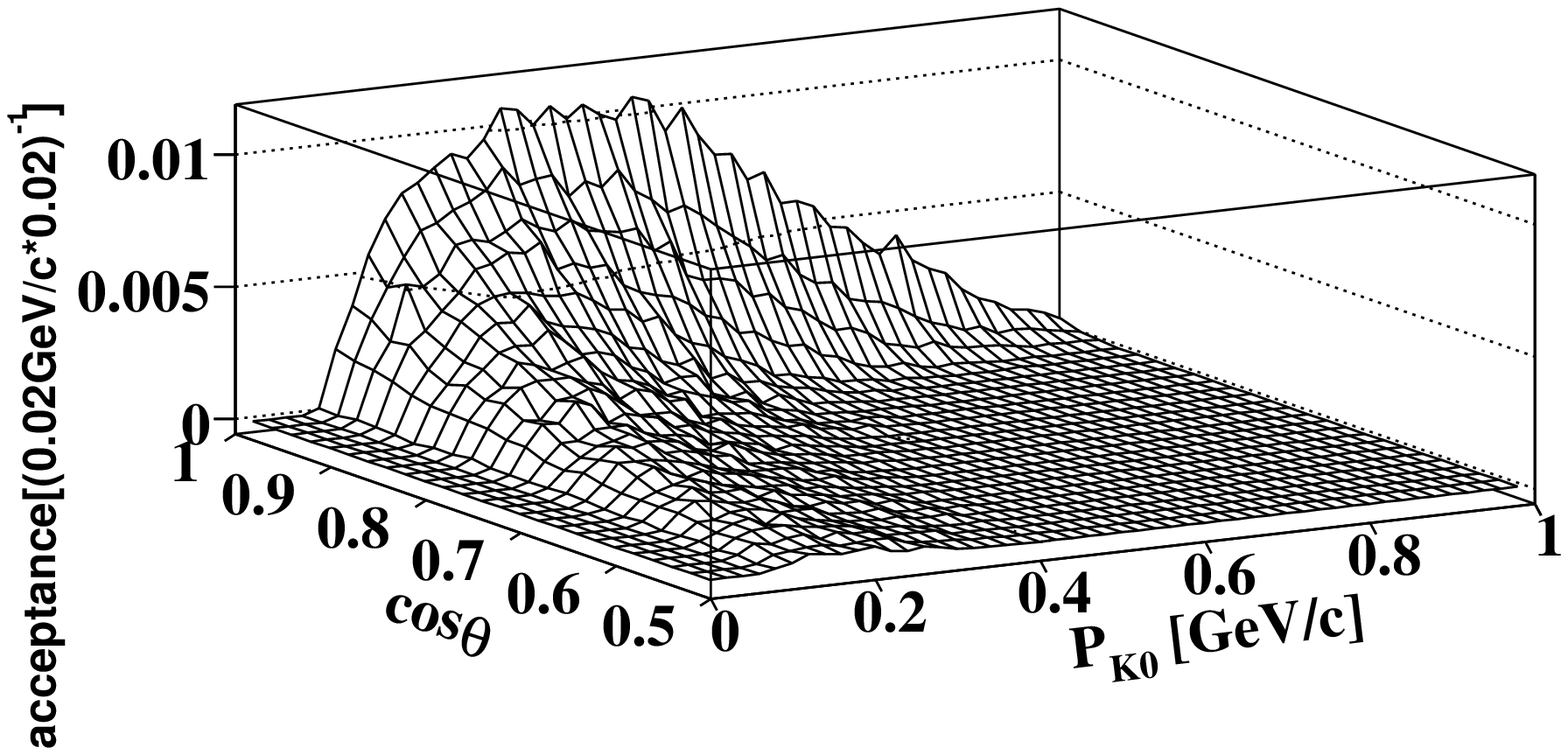}
 \includegraphics[width=8cm]{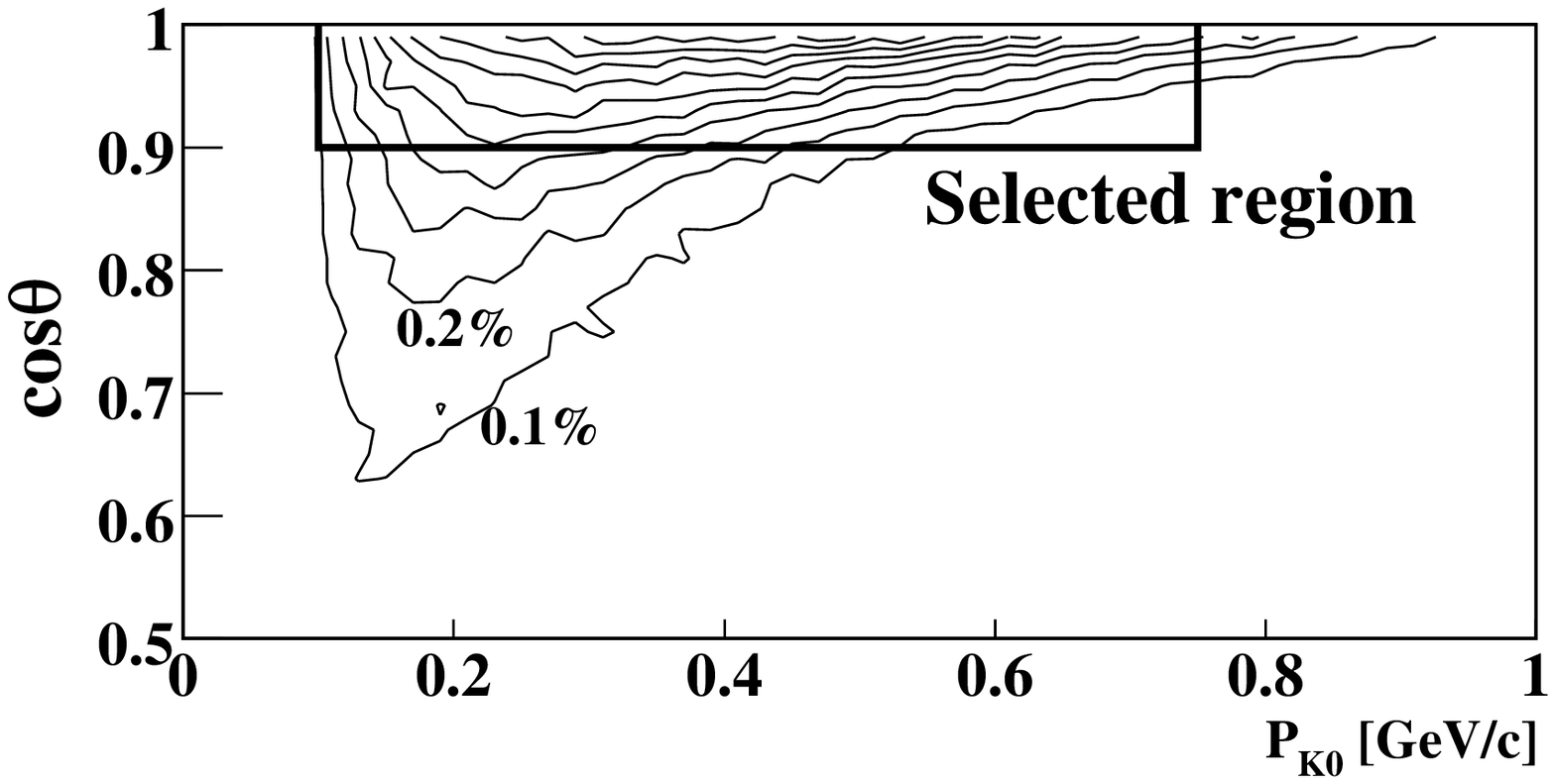}
 \caption{\label{Fig:acpt}
 Acceptance map of NKS evaluated by simulation shown
 as a function of
 the $K^0$ momentum and the cosine of the emitted angle with the photon beam line in the laboratory frame.
 The solid box in the right figure shows the accepted region, which has an efficiency larger than 0.1\%.
 }
\end{figure*}

The acceptance ($\epsilon_{acpt}(p,\cos\theta)$)
shown in Fig.~\ref{Fig:acpt}(left)
was evaluated 
by analyzing the simulated data
generated by the Geant4 program
taking into account the geometry of NKS,
the trigger condition
and detector resolutions.
As demonstrated in Fig.~\ref{Fig:acpt}, 
the acceptance reached a broad maximum in the forward region and momentum region of about 0.3 GeV/c.

In the following,
the data in the selected region, $0.1 < p_{K^0,Lab} < 0.75$ GeV and $0.9 < cos\theta_{K^0,Lab} < 1.0$ as indicated in Fig.~\ref{Fig:acpt}(right),
are analyzed.

\begin{figure*}
 \includegraphics[width=8cm]{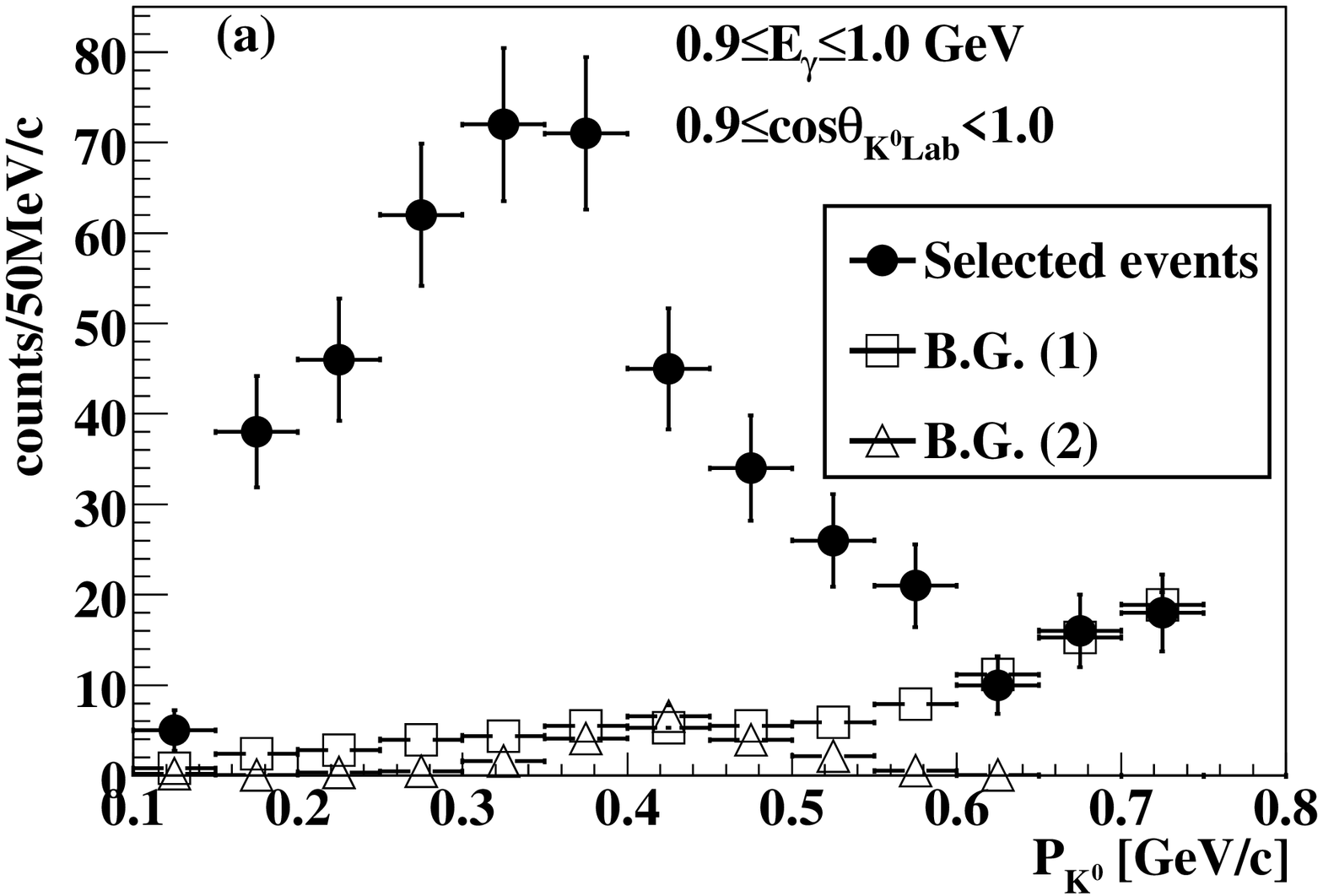}
 \includegraphics[width=8cm]{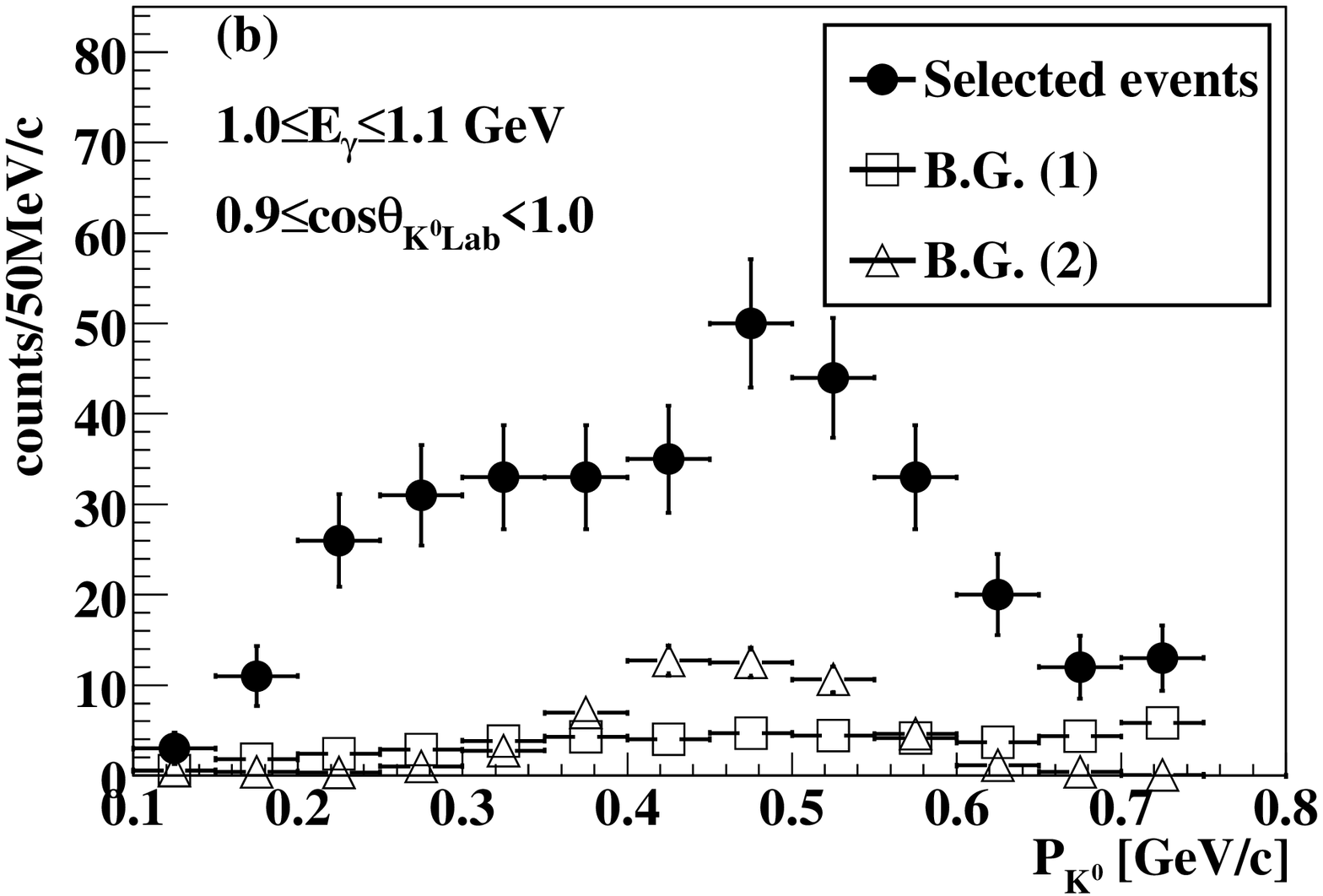}
 \caption{\label{Fig:momraw}
 Momentum distributions for $K^0$ in the selected region of NKS acceptance
 for events in the decay volume (closed circles), the background from the target region (open squares, B.G.~(1))
 and the combinatorial background (open triangles, B.G.~(2)).
 The error bars show the statistic errors
 and the fitting error for the scale factor of the backgrounds.
 }
\end{figure*}

Figure \ref{Fig:momraw} shows the $K^0$ momentum distribution
in the selected regions
for the two beam energies.
The contributions of the backgrounds described in Sect.~\ref{Sect:BG} 
are also overdrawn.

The number of tagged photons irradiating the target ($N'_\gamma$) was obtained 
to be
1.03$\times$10$^{10}$ for $0.8 \leq E_\gamma < 0.9$ GeV,
1.16$\times$10$^{10}$ for $0.9 \leq E_\gamma < 1.0$ GeV and
1.03$\times$10$^{10}$ for $1.0 \leq E_\gamma \leq 1.1$ GeV
from the scaler counts of tagging counters after correcting for the tagging efficiency and the analysis efficiency of the tagging counter array.
The tagging efficiency was estimated
using the data by a lead glass \v{C}erenkov counter (LG)
positioned 260 cm downstream from the experimental target position on the beam line.
The efficiency of tagger counter was totally derived to be 71\%.

The density of the deuterium target ($N_{target}$) was evaluated to be 0.173 g/cm$^3$
from its temperature and vapor pressure.
The expansion of the target cell due to the pressure of liquid deuterium
was estimated by a finite element method
and was found to be 1.58 mm for both sides.
Therefore, the typical area density was obtained to be 0.172$\times$10$^{24}$ [cm$^{-2}$].

The tracking efficiency ($\epsilon_{track}$) was estimated to be typically 81.4\%
using $\pi$ single-track events.
It has a small dependence on the emission angle of the track due to a high counting rate at the forward angles.
The efficiency decreases by 5\% at most in the very forward angle
and this dependence was taken into account in the analysis.

Table~\ref{SumEff} summarizes the efficiencies and their statistic and systematic errors.
\begin{table}[!h]
  \caption[Summary of efficiencies.]
  {Summary of efficiencies and numbers are listed.
  Some of them are not constant and listed for typical values.
  Systematic errors are estimated from the fluctuations among the experimental periods or the fluctuations of photon intensity.
  \label{SumEff}}
  \vspace{0.5em}
  \begin{tabular}{c|c|c|c}
   \hline\hline
                         & value                & statistical error & systematic error \\
   \hline\hline
   $\epsilon_{gate}$     & 95.4\%               & $<$ 0.1\%         & $\pm$ 4\%     \\ \hline
   $N'_\gamma (all)$     & 3.22$\times$10$^{10}$ & $\pm$ 1.0\%&    $\pm$ 4.7\%          \\ \hline
   $N_{target}$          & 0.172 [barn$^{-1}$]  & $<$ 0.1\%        & $\pm$ 0.6\%      \\ \hline
   $\epsilon_{DAQ}$      & 89.1\%               & $<$ 0.1\%        & $\pm$ 1.0\%       \\ \hline
   $\epsilon_{track}$    & 81.4\%              & $<$ 0.1\%       & $\pm$ 8.1\% \\ \hline
  \end{tabular}
\end{table}
The total systematic error was estimated to be about $\pm 10$\% by the square-root of the sum of the squares of the listed efficiencies.

\begin{figure*}
 \includegraphics[width=8cm]{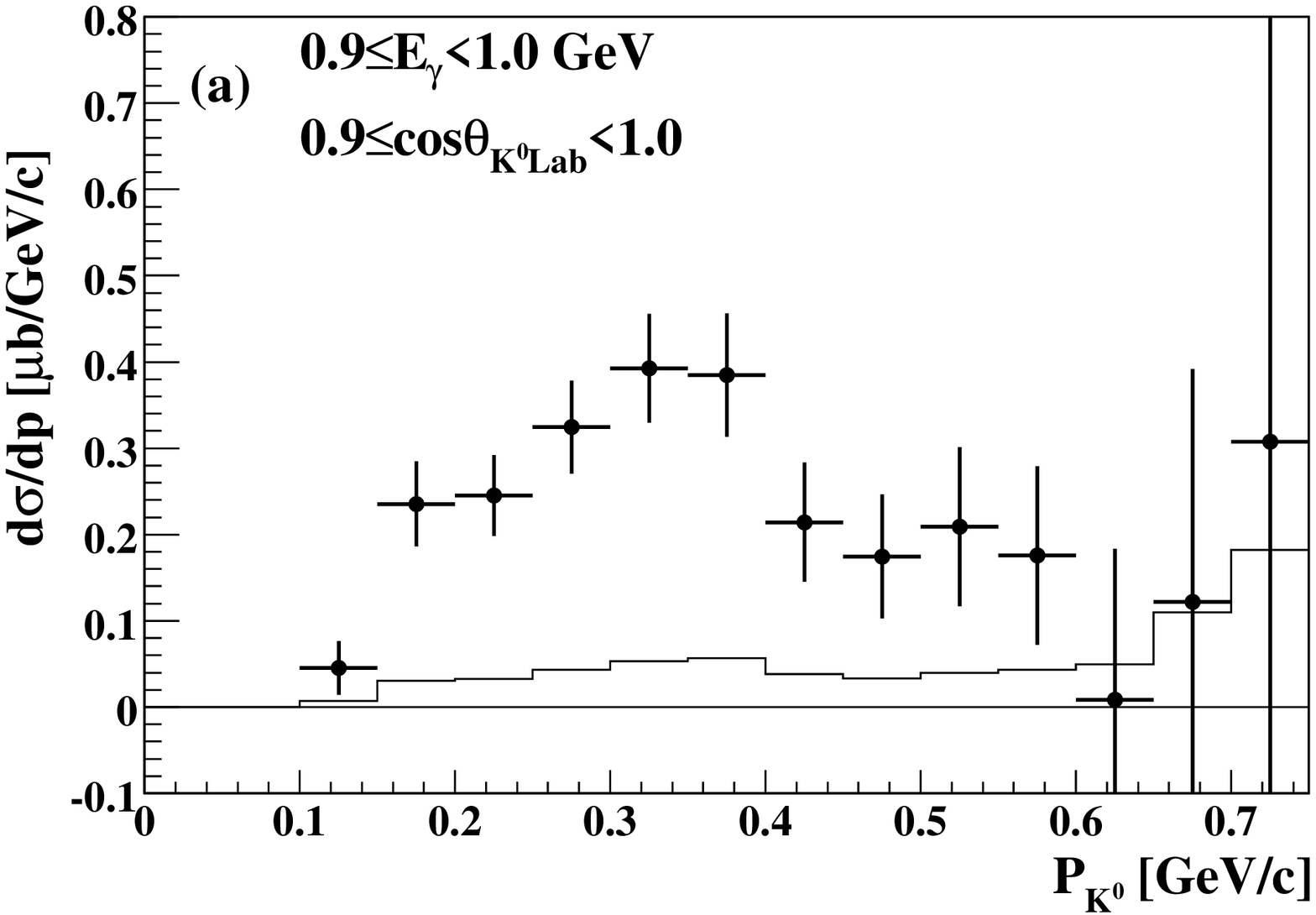}
 \includegraphics[width=8cm]{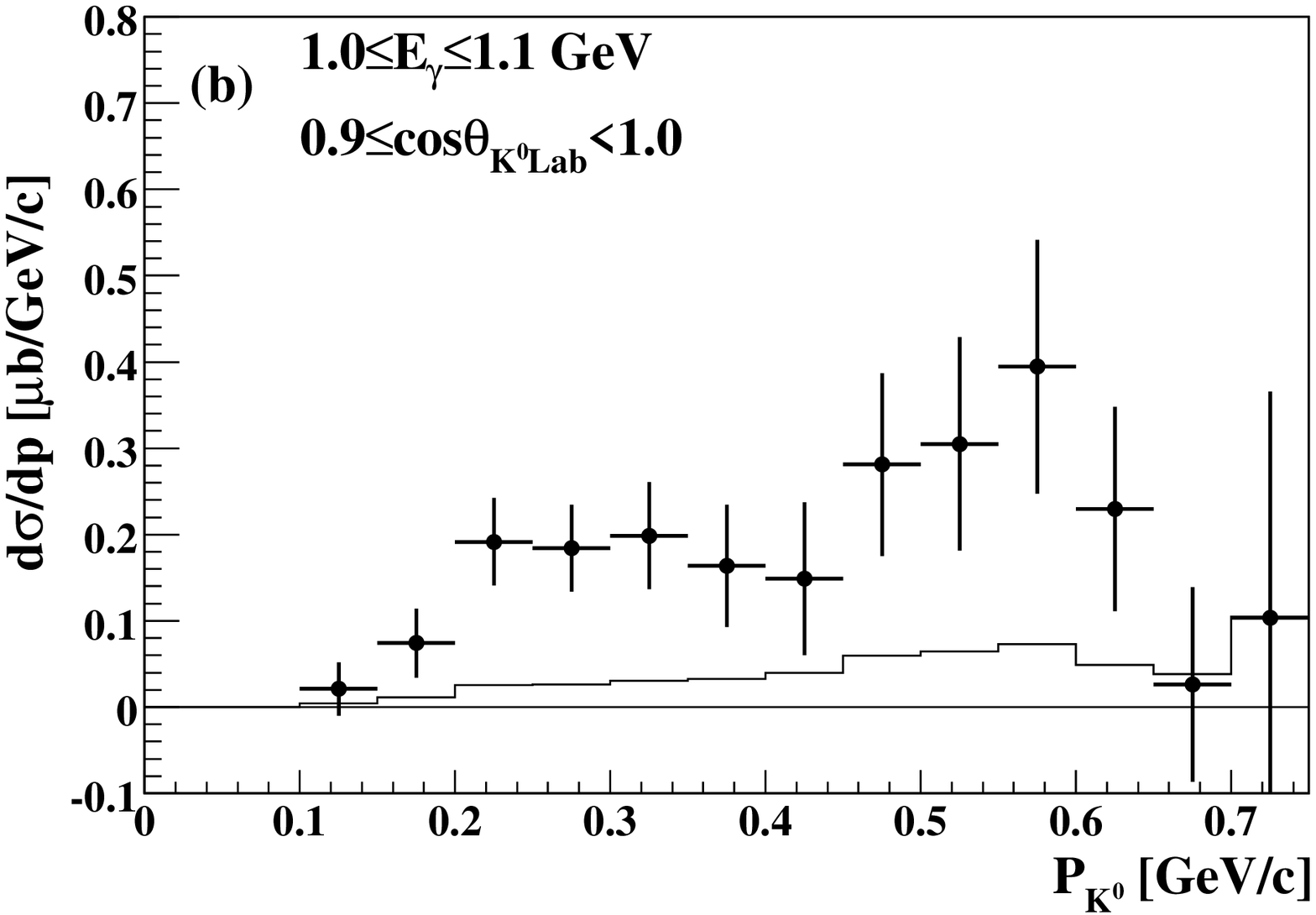}
 \caption{\label{Fig:mom}
 Inclusive momentum spectra for $K^0$ photoproduction in the selected region of NKS acceptance.
 The photon energies and kaon angles are displayed in the figure.
 The error bars show the statistical errors.
 The systematic error is represented by solid histograms on the base line.
 }
\end{figure*}

Taking into account all these efficiencies,
$K^0$ momentum spectra were derived in the two photon-energy regions as shown in Fig.~\ref{Fig:mom}.
The contributions of the background were subtracted in the cross sections.

\section{\label{Sect:result}Discussions}

\subsection{\label{Sect:TheoreticalPart}Theoretical calculations}
The inclusive cross sections for the $d(\gamma,K^0)YN'$ process,
where $Y=\Lambda$, $\Sigma^0$ and $\Sigma^+$, and $N'$ is the corresponding nucleon,
were compared with a simple spectator model calculation
in the plane wave impulse approximation
with a spectator nucleon ($N'$) \cite{Bydzovsky:2004ev}.
The energy of the target nucleon ($N$) is given by $E_N = m_d - E_{N'} = E_{K^0} + E_\Lambda - E_\gamma$
for the off-shell approximation,
to satisfy energy conservation in the elementary process,
and by $E_N = \sqrt{m^2_N + {\vec{p}_N}^{~2}}$ for the on-shell approximation.
The effective mass of the target nucleon is determined as 
$\tilde{m}_N = \sqrt{p^\mu p_\mu}$ and $\vec{p}_N = -\vec{p}_{N'}$.
The momentum distribution of the target nucleon is described by the non-relativistic Bonn deuteron wave function
OBEPQ (one boson exchange potential in q-space) \cite{Machleidt:1987hj}.

Differences in results caused by employing the on- and off-shell approximations and other deuteron wave functions were found to be small or negligible in the kinematical region studied in the present experiment \cite{Bydzovsky:2004ev}.
In the following analysis, we use the off-shell approximation.
Moreover, we assume that the contribution of the final-state interaction to the inclusive kaon momentum distribution \cite{Salam:2006kk}
is small in our kinematics
and therefore that our model is sufficient for a description of the process
(at least in the quasi-free region).

The largest uncertainty in the calculated $K^0$ momentum distribution came from the choice of the elementary amplitude \cite{Bydzovsky:2004ev}.
We used three different models for the elementary amplitude,
two isobar models, Kaon-MAID \cite{Lee:1999kd} and Saclay-Lyon A (SLA) \cite{Mizutani:1998sd},
and a simple phenomenological prescription,
to analyze the angular dependence of the elementary cross section in the center-of-mass frame (c.m.).

In the isobar models,
the cross sections for channels that include $K^0$ in the final state
were calculated assuming isospin symmetry for the strong coupling constants
and
appropriately adopting electromagnetic coupling constants
for the neutral mode by replacing those for the charged mode \cite{Lee:1999kd}.
For the t-channel meson resonances, 
the ratio of the coupling constants for neutral and charged modes is related to the ratio of the decay widths,
which is well known for $K$*(892) but unknown for the $K_1$ resonance \cite{Yao:2006px}.
The ratio for the latter, $r_{K_1 K\gamma}$, is therefore treated as a free parameter
and has to be determined from data.
In the case of the Kaon-MAID model,
the cross section of the $n(\gamma, K^0)\Lambda$ reaction can be predicted
since the parameter was determined by simultaneous fitting of the $K^0\Sigma^+$ channel.

On the other hand, in the SLA model,
the parameters were adjusted only from $K^+ \Lambda$ production data
and the model cannot predict the $K^0 \Sigma$ channel contributions.
If the value $r_{K_1 K\gamma} = -0.45$ determined in Kaon-MAID is used in SLA,
it gives a cross section several times larger than that of Kaon-MAID.
Therefore, when $K^0 \Lambda$ production is calculated with SLA,
$r_{K_1 K \gamma}$ is assumed as a free parameter.

In order to show the angular dependence of the elementary cross section 
preferred by
the data, a simple phenomenological parameterization in c.m., named PH, was also used:
\begin{eqnarray}
\frac{d\sigma}{d\Omega} = \sqrt{(s-s_0)}\left(1+e_0(s-s_0)\right)\left(a_0P_0(x)\right. \nonumber
\\ \left.+a_1P_1(x)+a_2P_2(x)\right) 
\label{Eq:ph}.
\end{eqnarray}
Here $P_l(x)$ are Legendre polynomials, $x = \cos(\theta_{K}^{c.m.})$,
$s$ is the square of the photon--nucleon c.m.~energy, and $s_0 = 2.603$ GeV$^2$, the $K^0\Lambda$ threshold.
Since the invariant cross-section is inversely proportional to the final c.m. momentum,
which behaves as $\sqrt{(s-s_0)}$ near the threshold,
the energy dependent factor of this form is
included
in (\ref{Eq:ph})
to regularize the invariant cross section.

\subsection{\label{Sect:ComparePart}Comparison of data with calculations}
\begin{figure*}
 \includegraphics[width=8cm]{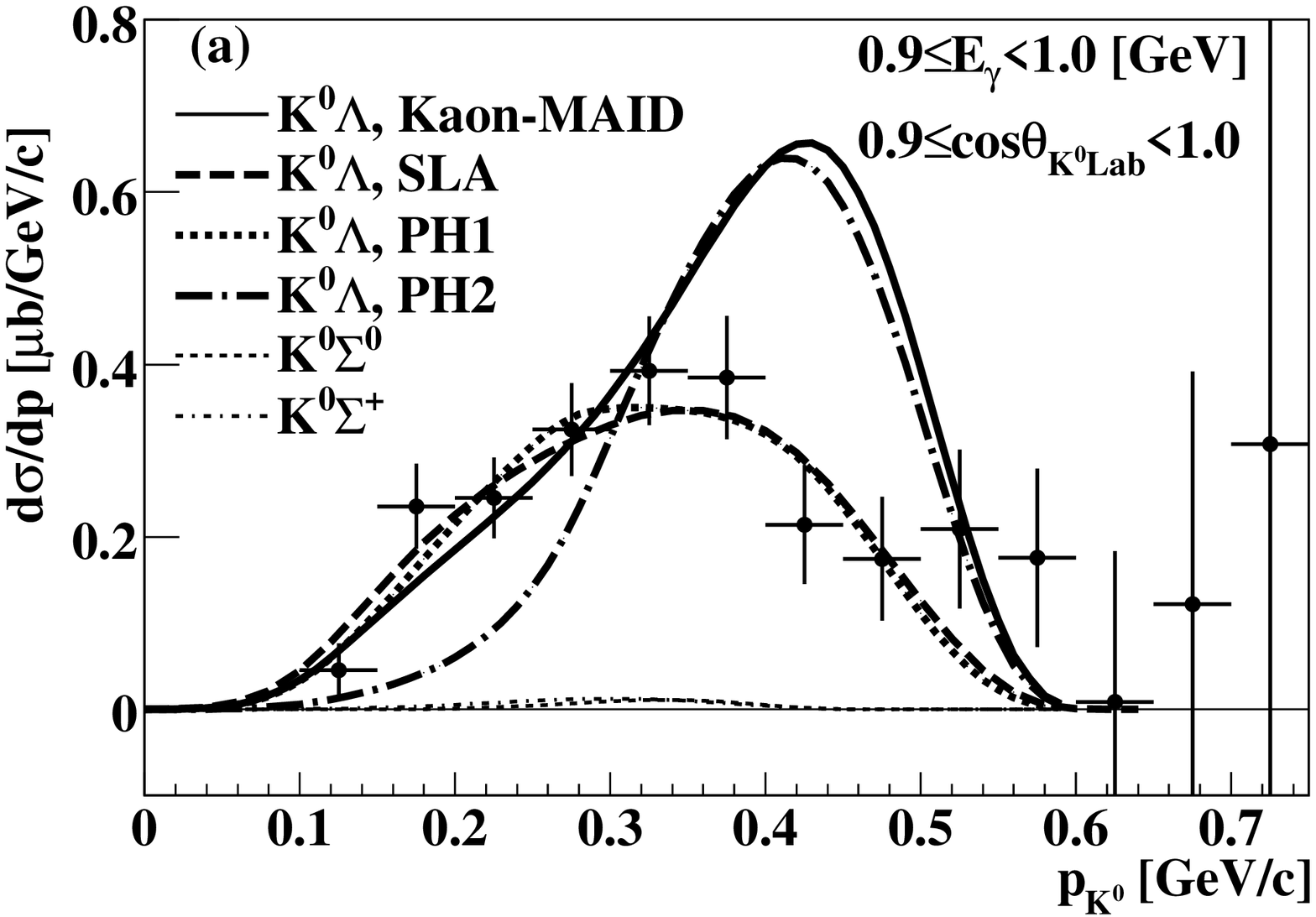}
 \includegraphics[width=8cm]{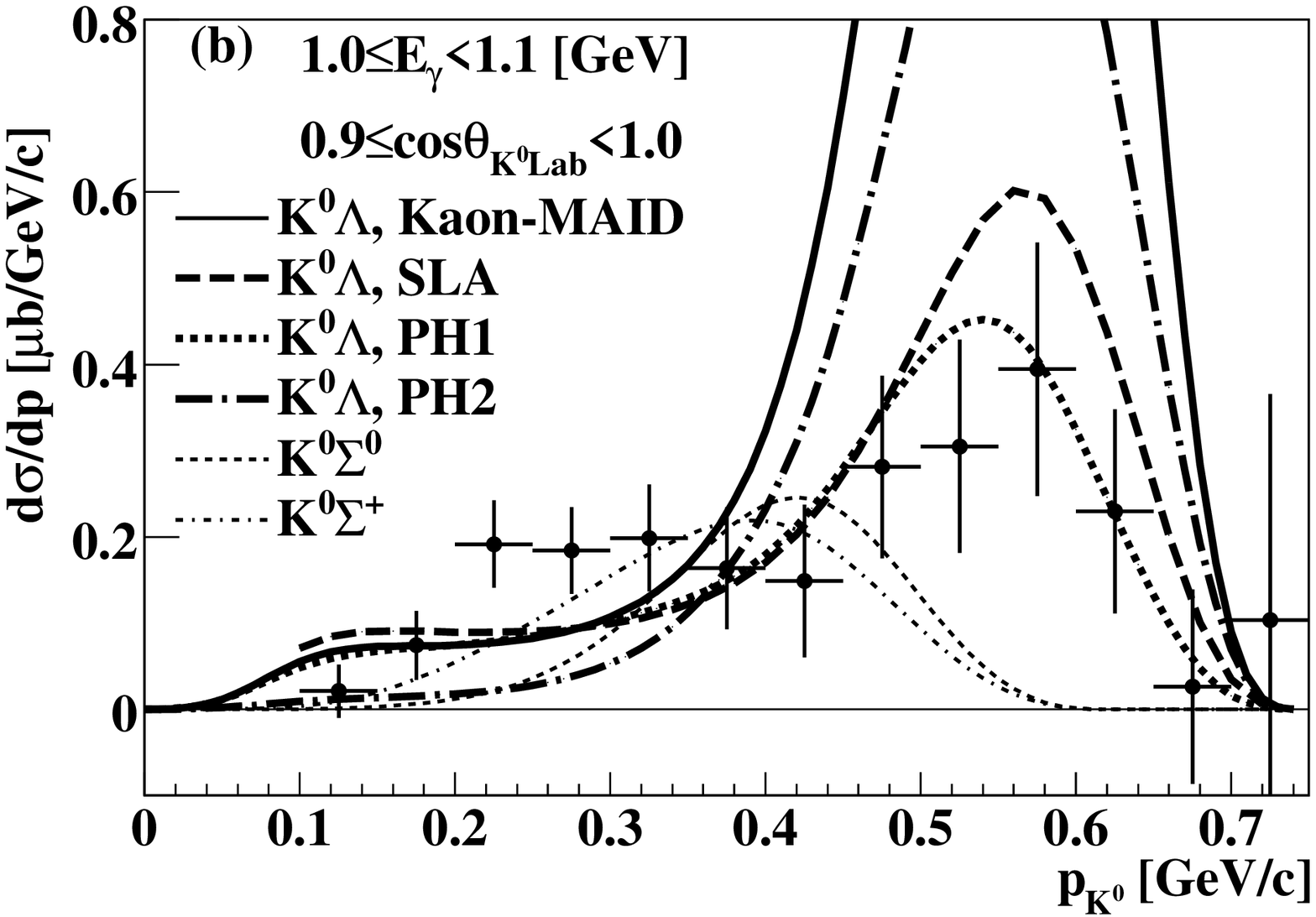}
 \caption{\label{Fig:momwt}
 Inclusive momentum spectra for $K^0$ photoproduction in the effective regions
 in comparison with calculations for the $K^0\Lambda$ channel using elementary amplitudes of Kaon-MAID \cite{Lee:1999kd} (solid line), SLA \cite{Mizutani:1998sd} with $r_{K_1 K\gamma}$ = $-$2.09 (dotted line), PH1 (dashed line) and PH2 (dash-dotted line) models.
 The contribution of the $K\Sigma$ channels calculated by using Kaon-MAID are also shown ( narrow dashed and narrow dash-dotted line).
 The photon energies and kaon angles are displayed in the figure.
 The error bars show statistical errors.
 The systematic error is not represented.
 }
\end{figure*}

The measured inclusive spectra of the $K^0Y$ channels are shown in Fig.~\ref{Fig:momwt}
together with theoretical curves.
Contributions of the $\Sigma$ photoproduction processes calculated with Kaon-MAID are also shown.
These contributions are estimated to be at the very most 5\% in the lower photon-energy region
but up to 60\% at the $K^0$ momentum of 0.3 GeV/c in the higher photon-energy region.
Therefore, the $r_{K_1 K\gamma}$ parameter of SLA was deduced from the data only in the lower photon-energy region
where the contribution of the $\Sigma$ channels is sufficiently small. 
The enhancement of the data points around 0.3 GeV/c seen in the higher photon-energy region
can be attributed to $\Sigma$ photoproduction, which however cannot be evaluated with SLA.

In both energy regions,
it is found that Kaon-MAID gives cross sections larger than the present measured spectra
and the shape of the calculated momentum spectra are considerably different from that of the data. 

The theoretical spectrum with SLA was fitted with the $r_{K_1K\gamma}$ parameter 
in the lower photon-energy region.
The magnitude and shape of the momentum spectrum are reproduced well for $r_{K_1K\gamma} = -2.087$.
Note that this value is related to the ratio of the decay width of the $K_1$ meson,
$\Gamma_{K_1^0 \rightarrow K^0\gamma}/\Gamma_{K_1^+\rightarrow K^+\gamma} = r_{K_1K\gamma}^2 = 4.36$.
The magnitude of the cross section in the higher energy region is reproduced better by SLA than Kaon-MAID.

\begin{figure*}
 \includegraphics[width=8cm]{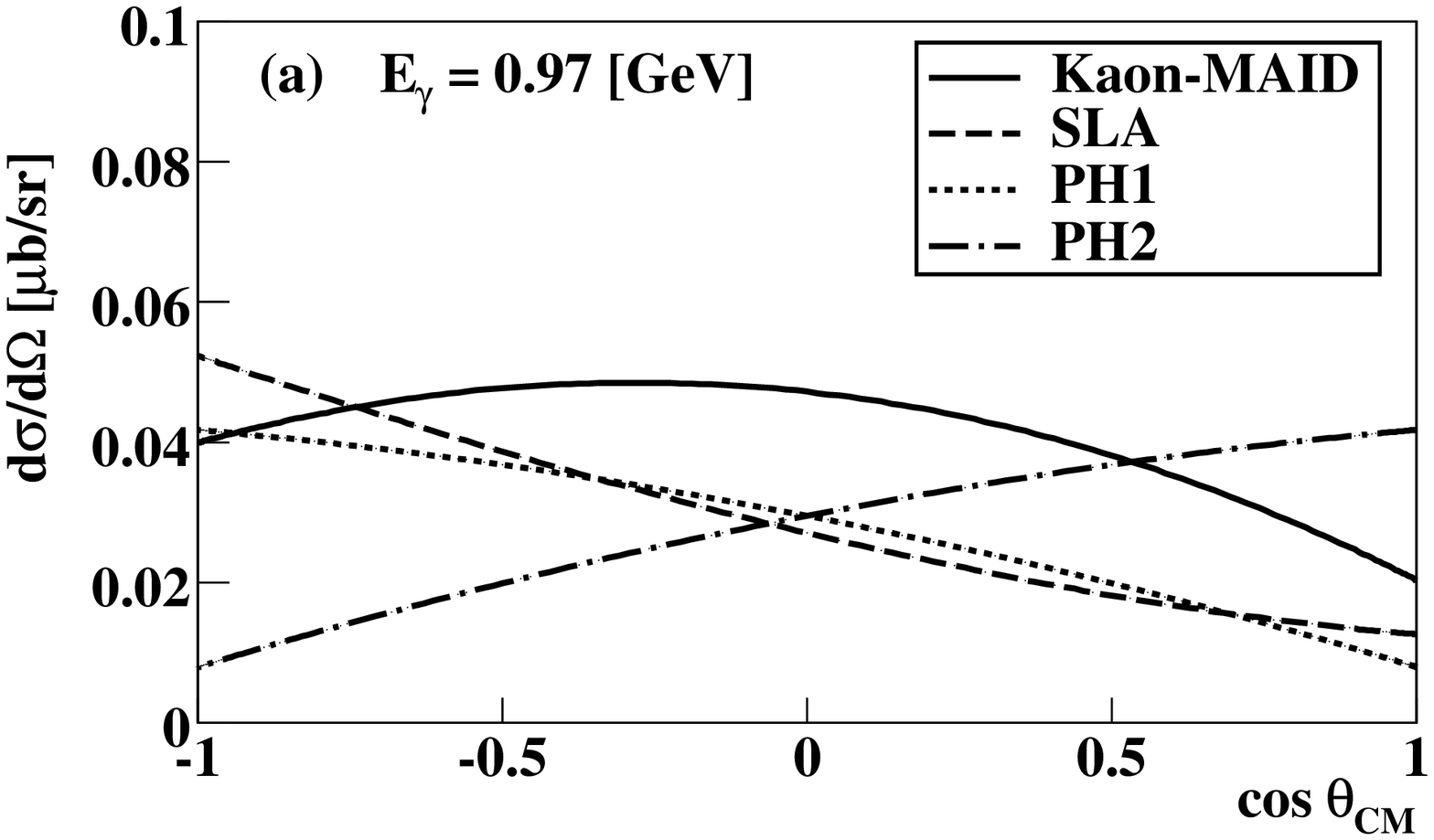}
 \includegraphics[width=8cm]{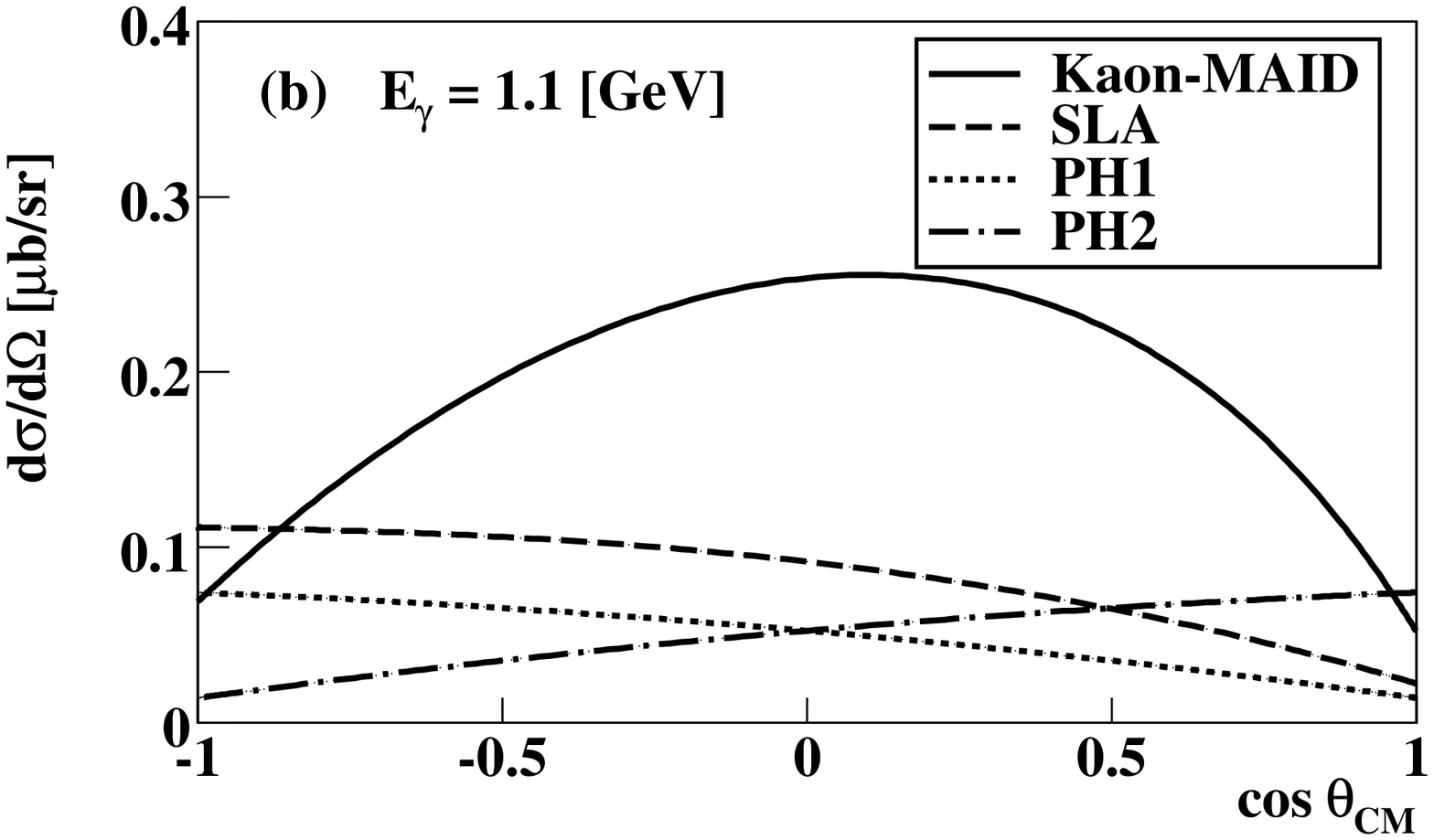}
 \caption{\label{Fig:elemC}
 Angular distributions of the elementary cross section of $K^0\Lambda$ production
 in the c.m.~system for (a) $E_\gamma=0.97$ GeV and (b) $E_\gamma=1.1$ GeV.
 The Kaon-MAID \cite{Lee:1999kd} (solid line), SLA \cite{Mizutani:1998sd} with $r_{K_1 K\gamma}$. = $-$2.09 (dotted line), PH1 (dashed line) and PH2 (dash-dotted line) models are shown.
 }
\end{figure*}

The difference of the shape of the momentum spectra in the laboratory system
between Kaon-MAID and SLA in the lower photon-energy region is caused
by the different angular distribution of the elementary cross sections in the c.m.~system,
as shown in Fig.~\ref{Fig:elemC}.
The data, therefore, suggest that an enhancement of the elementary cross section in the backward hemisphere is crucial
to explain the $K^0$ momentum spectrum shape.
On the other hand, in the higher photon-energy region,
the difference of the momentum spectra
is due not only to the angular distribution but also to the magnitude of the cross section, as seen in Fig.~\ref{Fig:elemC}(b).

To confirm these points,
the spectrum in the lower photon-energy region were fitted
by Eq.~\ref{Eq:ph},
in which the angular distribution is given by Legendre polynomials up to second order.
The best fit parameters (PH1) were obtained:
$a_0=0.0884$, $a_1=-0.0535$, $a_2=-0.0098$ and $e_0=-0.132$ with $\chi^2/n.d.f.=1.19$.
The phenomenological parametrization with the opposite sign of $a_1$ (PH2), giving the inverse angular distribution
to that of PH1 is also shown in Fig.~\ref{Fig:momwt} and \ref{Fig:elemC}.
In Fig.~\ref{Fig:momwt}, it is shown that PH1, having a backward angular distribution in c.m. similar to that of SLA, gives a very good agreement with the present data in both photon-energy regions.
On the other hand, PH2, giving a momentum spectrum similar to that of Kaon-MAID and a forward peaked angular distribution in contrast to PH1, clearly contradicts the present result.
Thus, a comparison of the data and predictions of models favors 
an angular distribution of the $n(\gamma,K^0)\Lambda$ reaction
in the c.m.~system that is peaked at backward angles.

\begin{figure*}
 \includegraphics[width=10cm]{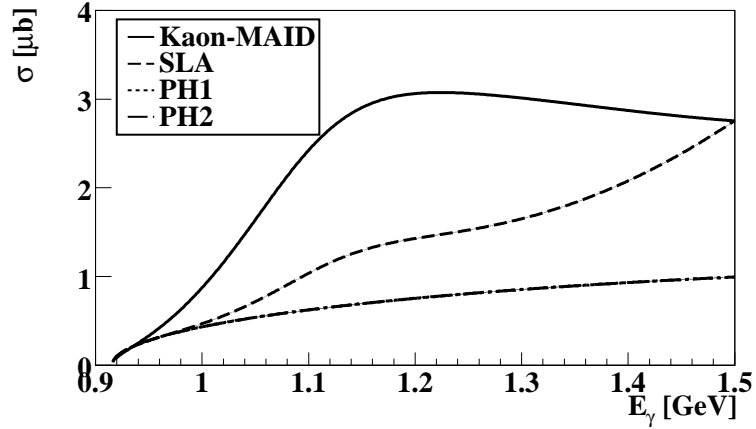}
 \caption{\label{Fig:elemE}
 Calculated photon-energy dependence of the total cross section of $K^0\Lambda$ production
 for the Kaon-MAID \cite{Lee:1999kd} (solid line), SLA \cite{Mizutani:1998sd} with $r_{K_1 K\gamma}$. = $-$2.09 (dotted line), PH1 (dashed line) and PH2 (dash-dotted line) models are shown.
 }
\end{figure*}

In Fig.~\ref{Fig:elemE}, the photon energy dependence of the total cross section for $K^0\Lambda$ photoproduction is given for SLA with $r_{K_1K\gamma} = -2.087$, Kaon-MAID and the PH1 and PH2 parameterizations.
The $K^0$ total cross sections are not obtained in the present experiment
due to the limited kinematical acceptance for $K^0$.
 Kaon-MAID predicts a very sharp rise in the photon-energy dependence of the $K^0$ total cross section near the threshold ($E_\gamma < 1.2$ GeV),
while SLA, as well as PH1 and PH2, gives a flatter energy dependence and much smaller cross section in the threshold region.


\section{\label{Sect:sum}Summary}

$K^0$ photoproduction on a deuterium target was studied in the threshold region for the first time.
The experiment was conducted at LNS of Tohoku University following the measurement of quasi-free photoproduction of $K^0$ on a carbon target.

Inclusive momentum distributions were obtained and compared with theoretical calculations.
The calculations were performed in the plane wave impulse approximation
using a realistic Bonn wave function OBEPQ for a deuteron,
firstly with the recent isobar models, Kaon-MAID and Saclay-Lyon A,
and secondly with simple phenomenological parameterization.
Based on the comparison,
the angular distributions of the elementary cross sections were examined.

The best agreement with the data was achieved with the SLA model, in which the ratio of the neutral-to-charged electromagnetic coupling constants for the $K_1$ meson was 
obtained to be $r_{K_1K\gamma}$ = $-$2.09 by fitting data in the photon energy region, $E_\gamma = 0.9$--1.0 GeV,
and with phenomenological parametrization fitted
in the photon energy region, $E_\gamma = 0.9$--1.0 GeV.
Both models show enhancement of the c.m. cross section in the backward hemisphere.
The present result, therefore, sets constraints on models for the photo- and electroproduction of kaons in the threshold region.
This study shows that the $n(\gamma,K^0)\Lambda$ reaction with its unique feature is important for the investigation of the strangeness photoproduction.

\section{\label{Sect:ack}Acknowledgment}
This work is supported by Grants-In-Aid for Scientific Research from The Ministry of Education of Japan, Nos. 09304028, 12002001, and 14740150
and by the Grant Agency of the Czech Republic under grant No. 202/05/2142.


\end{document}